\documentclass[12pt]{iopart}

\usepackage{iopams}
\usepackage{graphicx}  
\usepackage{color}

\begin{document}

\title{Master equation for cascade quantum channels:  a collisional approach}
\author{V. Giovannetti$^{1}$ and G. M. Palma$^{2}$}

\address{$^1$ NEST, Scuola Normale Superiore and Istituto Nanoscienze-CNR, 
piazza dei Cavalieri 7, I-56126 Pisa, Italy \\ $^2$NEST  Istituto Nanoscienze-CNR and Dipartimento di Fisica,
Universita' degli Studi di Palermo, via Archirafi 36, I-90123 Palermo, Italy}

\begin{abstract}

It has been recently shown that collisional models can be used to derive 
a general form for the master equations which describe the reduced time evolution of a composite multipartite quantum system, 
whose components   ``propagate"
in an environmental medium which induces correlations among them via a cascade mechanism. 
Here we analyze the fundamental assumptions of this approach showing how some of them can be lifted when passing into a proper
interaction picture representation.

\end{abstract}

\maketitle

\section{Introduction} 
\begin{figure}[t]
\begin{center}
\includegraphics[width=400pt]{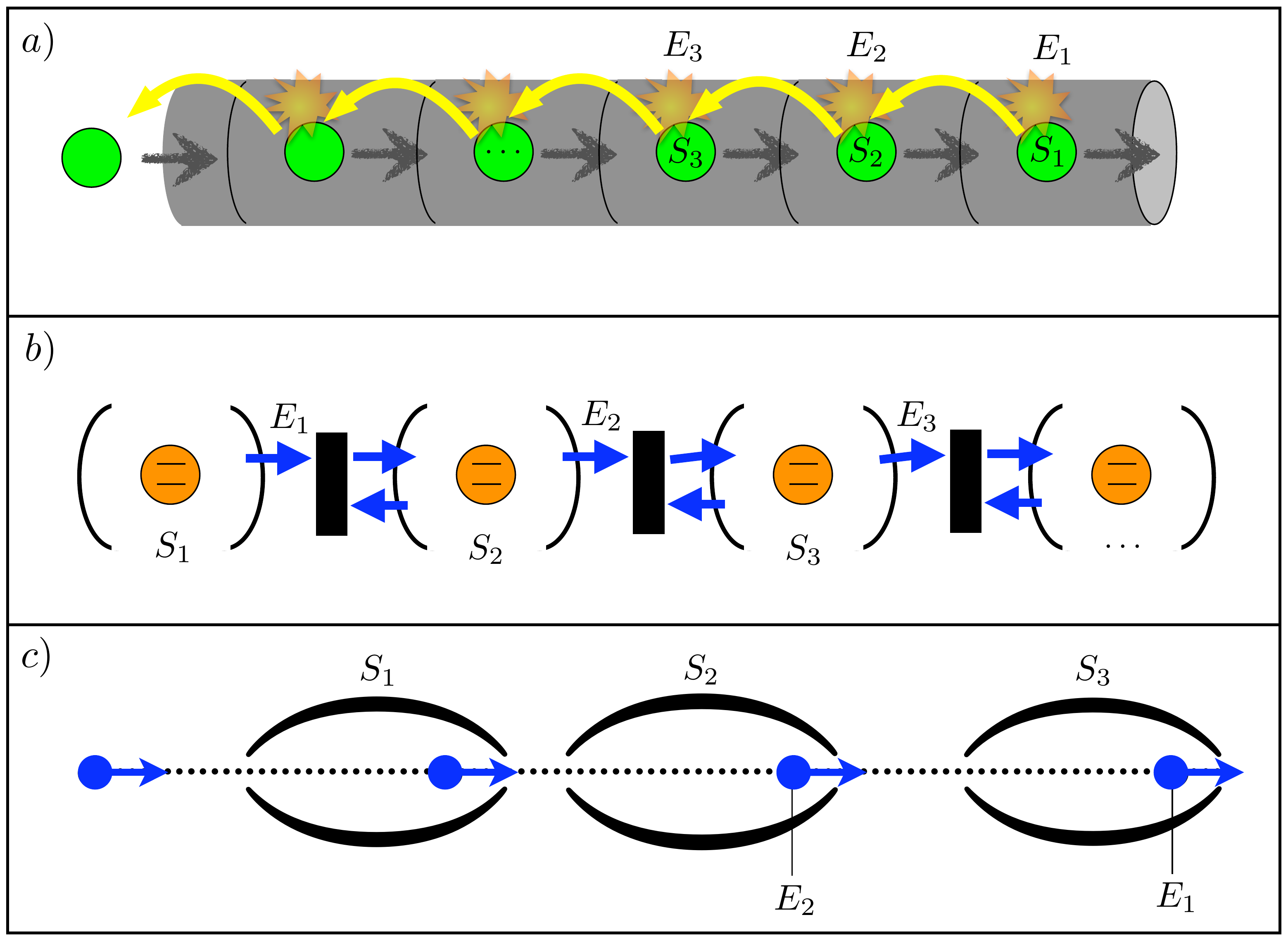} 
\caption{(Color online) Prototypical examples of quantum systems which admit a representation in terms of cascade quantum master equations.
 a) A string of quantum information carriers (say the photon pulses represented by the green element of the figure) propagating along a fiber while losing photons along the way.
 Here the sub-environment $E_1, E_2, \cdots$ are associated with different sectors of the fiber
 which effectively describe the transmission line within a lumped model approach.
 If the propagation speed of the pulses is sufficiently high, the excitations they waste in a given
 sector of the fiber cannot be re-absorbed by the same pulse: still the subsequent pulses have a certain probability of  absorbing it. The net result is an indirect, unidirectional coupling between
 the pulses mediated by the fiber which allow quantum signals to pass from one carrier to the subsequent ones as schematically shown by the yellow arrows of the figure. 
 b) A set of atoms or ions (represented by the orange elements in the figure) trapped into a series of QED cavities which exchange photons (blue arrows), via unidirectional couplers (black elements)~\cite{G,G1}.
 c)  An array of cavities $S_1, S_2, \cdots$ crossed sequentially by atoms $E_1, E_2,\cdots$ (blue elements) of a beam. The injection rate is such that the atoms cross the cavities one by one. The atoms are initially prepared all in the same state. }
\label{fig3}
\end{center}
\end{figure}

In the study of  the open dynamics of a multipartite quantum system  it is frequently made the simplifying assumption that each subsystem interacts with its own local environment. 
In the language of quantum communication~\cite{BEN&SHOR}
 this is equivalent to saying that the resulting time evolution is memoryless i.e. that  the noise tampering the communication  acts independently on each local component (information carrier) of the transmitted quantum message. 
 In recent years, however,
it has been shown that interesting new features emerge when one makes the realistic assumption that the action of a channel over consecutive uses is  correlated~\cite{MPV, BM,KW,VJP,PV,DARRIGO,LUPO}. 
Such  correlations have been  phenomenologically described in terms of a Markov chain which gives the joint 
  probability distribution of the local Kraus operators acting on the individual carriers~\cite{MPV}. 
  Alternatively they have been effectively represented in terms of local interactions of the carriers with a common multipartite environment  
  which is originally prepared into a correlated (possibly entangled) initial state~\cite{PV}, or 
with a structured environment composed by local and global components~\cite{BM,KW,VJP}.
These models, although physically well motivated do not have an intrinsic time structure, in other words they are unable to describe a situation in which the information carriers interact one after the other with an environment which in the meanwhile evolves.
For instance consider the case in which an ordered sequence of  spatially separated 
 photon pulses carrying information in their photon number, 
 propagates at constant speed in a non passive, lossy optical fiber 
characterized by (relatively) slow reaction times.  
 If the speed of the pulses is sufficiently high,  one might expect that, thanks to 
the mediation of the fiber,  excitations from one pulse could be passed to the next one modifying their internal states via a (partially incoherent)
cascade mechanism (see Fig.~\ref{fig3} a) for a schematic representation of the process). The net result of course is the creation of delocalized excitations over the whole string of carriers as they proceed along the fiber (the extent of such delocalization depending upon the ratio 
between the transmission rate at which the carriers are fed into the fiber and the dissipation rate of the latter).
Alternatively, consider the case in which a linear array of local quantum systems (say a set of  driven QED cavities as in Fig.~\ref{fig3} b), or a set of quantum dots composing a quantum cascade laser~\cite{QCL}), are indirectly 
coupled via  unidirectional environmental mediators (the photons emitted by the cavities or by the dots) 
which passing from one system to the other, allow them to exchange excitations
~\cite{G,G1,EXAMPLE00,EXAMPLE01,EXAMPLE0,EXAMPLES1,EXAMPLES2}. As in the previous case, the formation of 
delocalized excitations is expected as time passes (in this case the delocalization of the excitation
will depend upon the product between  the damping rate of the mediator  and the distance between two consecutive quantum systems). 

 The general form for  master equations which describe these situations has been recently derived in Ref.~\cite{PRL}
by adopting a collisional approach~\cite{SZS,ZSB} to describe the system/environment coupling.
We aim to review these findings, focusing on some technical aspects of the problem
which allows us to lift some of the assumptions of Ref.~\cite{PRL}.
The manuscript is organized as follows. In Sec.~\ref{sec:collisional} we describe the collisional model, its continuous time limit (Sec.~\ref{sec:continuous}) 
and the basics properties of the associated master equation for cascade quantum systems. 
In Sec.~\ref{stabilitycond} we then pass to discuss the fundamental assumption which underline the derivation (namely 
the environment stability condition under collisions). Here we first show how local free evolution term can be embedded in the derivation (Sec.~\ref{sec:includ}). Then 
 we prove that the stability condition can always be enforced by passing through an interaction picture representation 
 which defines a more  ``stable" effective coupling with  the environment (Sec.~\ref{sec:enforcing}). Conclusions and final remarks are presented in Sec.~\ref{conc}.

\begin{figure}[t]
\begin{center}
\includegraphics[width=300pt]{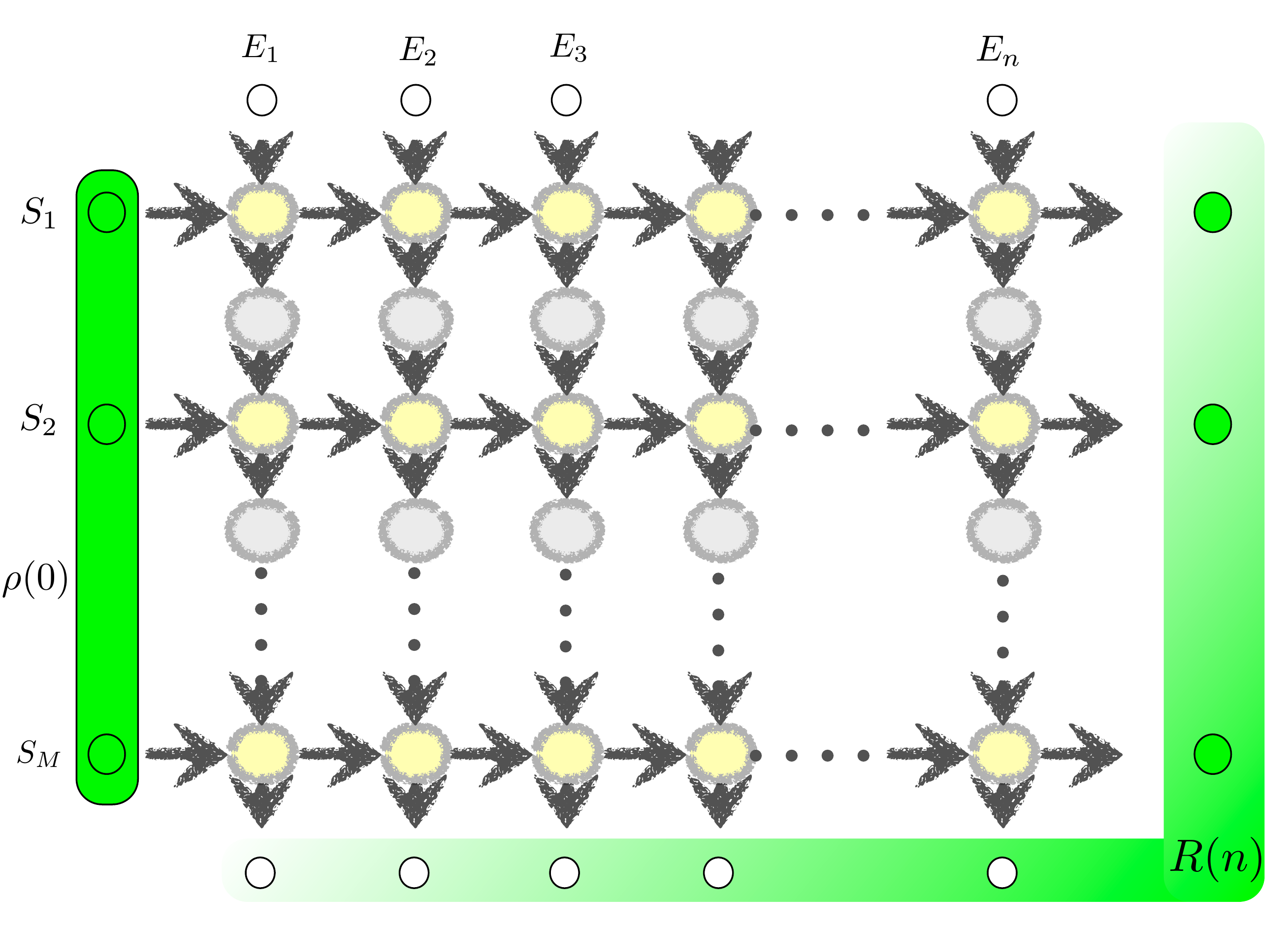} 
\caption{(Color online) Schematic of the collisional model. The horizontal lines describe an ordered set of  carriers  $S_1, S_2, \cdots$ which interact with an ordered set of (possibly infinite) identical local sub-environments  $E_1, E_2, \cdots$ via local unitaries $\mathcal{U}_{S_mE_n}$ represented by the yellow elements ($\eta$ being the initial state of the $E_j$s). Between collisions each sub-environment evolves according to the  ``damper" maps $\mathcal{M}$ (represented in the figure by the grey elements). The overall dynamics can be described as a ordered sequence of row or of column super-operators -- see Fig.~\ref{fig2} for details.}
\label{fig1}
\end{center}
\end{figure}
\section{The collisional model} \label{sec:collisional}

Consider a multipartite quantum system ${\cal S}$, composed by $M$ - not necessarily identical - ordered subsystem $S_1,  S_2 \cdots S_M$ which we shall refer to as  the information carriers of the model. 
They  are assumed to be prepared in a possibly entangled initial state $\rho(0)$ and to evolve in time due 
to the interactions with a multipartite  environment ${\cal E}$ consisting of a collection of sub-environments 
$E_1, E_2, \cdots$. Following the collision model of irreversible dynamics presented in Refs.~\cite{SZS,ZSB}, the carriers/sub-enviroment couplings 
 are described via a sequence of pairwise, time-ordered unitary interactions  which in our case are organized as in the scheme shown in~Fig.~\ref{fig1}. 
 According to it, each element
 of ${\cal S}$ interacts with all the elements of ${\cal E}$ in such a way that given $m'\geqslant m$ and $n'\geqslant n$
integers,  the  ``collision" between $S_m$ and $E_n$ is assumed to happen {\em before} the one involving $S_{m'}$ and $E_{n'}$. In particular, this implies that   the $S_1, E_1$ interaction   takes place before the couplings between $S_1,E_2$, the coupling between $S_2,E_1$, and  the coupling between $S_2,E_2$. Similarly the $S_2, E_2$ coupling is assumed to   come
after the  $S_2,E_1$ and the $S_1,E_2$ couplings, while  no specific
ordering is imposed on these last two events. 
Within this  theoretical framework the temporal evolution of  $m$-th carrier $S_m$ can be then be described 
trough  the action of the  following joint unitary evolution 
\begin{eqnarray}\label{hh2}
U_{S_m {\cal E}}^{(n)} 
 :=  U_{S_m,E_n}\;U_{S_m,E_{n-1}} \;\cdots \; U_{S_m,E_2} \; U_{S_m,E_1}\;,
\end{eqnarray} 
where, for instance, 
\begin{eqnarray}
U_{S_m,E_n} := \exp[ -i g H_{S_mE_n} \Delta t] \;,
\end{eqnarray}
 is the transformation that characterizes the ``collision"  between $S_m$ and $E_m$. 
In this expression $\Delta t$ is the collision time, $g$ is an intensity parameter that gauges the 
strength of the coupling, while $ H_{S_mE_n}$ is the coupling Hamiltonian which, without loss of generality, we write as
\begin{eqnarray}\label{ggrss}
 H_{S_mE_n} := \sum_\ell \; A^{(\ell)}_{S_m}  \otimes B^{(\ell)}_{E_n} \;,
 \end{eqnarray} 
 with $A^{(\ell)}_{S}, B^{(\ell)}_{E}$ being Hermitian operators.    It is worth stressing that in writing Eq.~(\ref{hh2}) one implicitly assumes that 
 a given carrier never interact twice with the \emph{same} sub-enviroment. This hypothesis is typically enforced in collisional models  which aim to describe 
Markovian processes -- see however Ref.~\cite{BP} for an alternative approach. Its validity relies  on the existence of a two well separated time scales: a fast one, which defines the typical correlation times of the
environment, and a slow one, which instead defines the dissipative effects on the system of interest (i.e. the carriers) induced by the coupling with the bath. 
Such assumption of course is not always fulfilled and when enforced it inhibits the possibility of feedback mechanisms where the state of the system at a given time is influenced 
by the entire evolution history. Notice however that in the scenario we are considering here,  the global Markovian structure of the coupling~(\ref{hh2}) 
doesn't prevent the possibility that different carriers could have a non trivial causal influence on each other. In other words, as schematically shown in Fig.~\ref{fig3} a), 
 quantum ``information" can be transferred from one carrier to the other through the intermediation of the environment.

In top of  the processes  described by the unitary couplings~(\ref{hh2}) 
we also assume that  between two consecutive collisions each sub-environment evolves according to the action of  a local Completely Positive 
(CP) map $\mathcal{M}$.
The latter is introduced to effectively account for the internal dynamics of ${\cal E}$: in particular  the transformations $\mathcal{M}$ 
mimics the relaxation processes that may take place within the environment alone (e.g. originating from the mutual interactions between its various parts) and which in principle involve timescales different from those that define the {\em rate} of the collisional events.
In other words, as in Ref.~\cite{VJP}, the mappings ${\cal M}$  act as ``dampers" for the information that percolates 
from one carrier to the subsequent one\footnote{In what follows we will work under the simplifying assumption that the {\em same} CP transformation 
acts among any two collisions -- the generalization to the case in which the ${\cal M}$ change passing from one collisional event  to the other being straightforward.}: how effective such damping is, it depends of course
upon the rate at which two subsequent carriers  approach the same sub-environemnt (i.e.  in the example
of Fig.~\ref{fig3} a), it depends upon the propagation velocity of the pulses along the fiber). Putting all together the resulting temporal evolution can hence be expressed in a compact form by 
observing that after the interactions with the first $n$ elements of ${\cal E}$ the global state $R(n)$ 
of the system and of the environment is obtained from the initial state $\rho(0) \otimes\eta^{\otimes n}$ as
\begin{eqnarray} 
R(n) = {\cal W}^{(n,M)} (\rho(0) \otimes\eta^{\otimes n})\label{erren}\;,
\end{eqnarray}
where ${\cal W}^{(n,M)}$ is the super-operator which describes
the collisions and the free evolutions of ${\cal E}$
 while
$\eta$ is the density matrix which describes the initial state of the sub-environments (for simplicity we assumed that all the $E_n$ are characterized by the same
initial state).
\begin{figure}[t]
\begin{center}
\includegraphics[width=400pt]{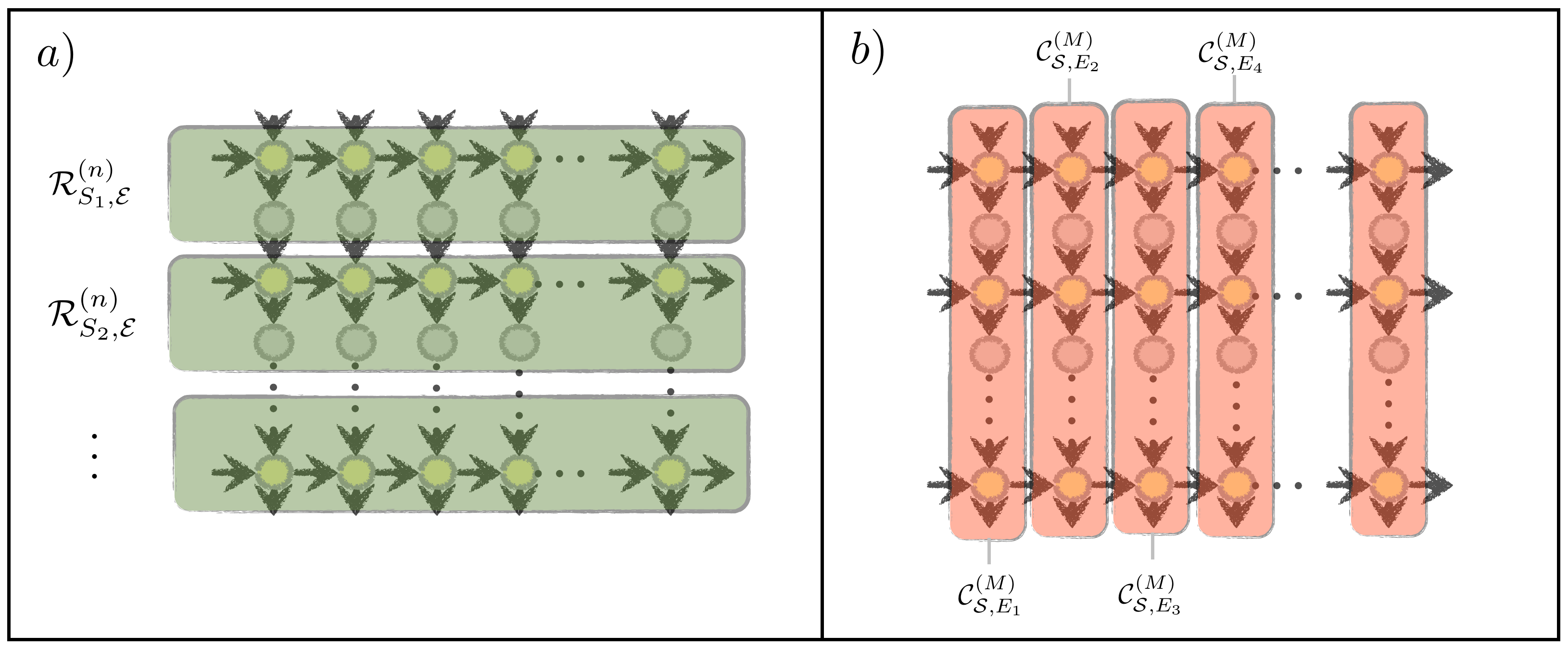} 
\caption{(Color online) Schematic of the decomposition of the evolution map ${\cal W}^{(n,M)}$ collisional model in terms of a) row operators as detailed in Eq.~(\ref{ROW}) 
 (in the figure these operators  are represented by the gray elements), or  b) column operators  (represented by the red elements) as detailed in Eq.~(\ref{colu}).}
\label{fig2}
\end{center}
\end{figure}
The ${\cal W}^{(n,M)}$ can be expressed as a composition of {\em row} super-operators 
stack in series one on top of the other  (see Fig.~\ref{fig2})
\begin{eqnarray}\label{ROW}
{\cal W}^{(n,M)}  =  {\cal R}^{(n)}_{S_M, {\cal E}} \circ  {\cal R}^{(n)}_{S_{M-1}, {\cal E}} \circ  \cdots\circ   {\cal R}^{(n)}_{S_2, {\cal E}} \circ  {\cal R}^{(n)}_{S_1, {\cal E}} \;, 
\end{eqnarray}
where we use the symbol ``$\circ$"  to represent the composition of super-operators 
and where 
\begin{eqnarray}
{\cal R}^{(n)}_{S_m,{\cal E}} := {\cal M}^{\otimes n} \;\circ\; {\cal U}_{S_{m},{\cal E}}^{(n)} \;.
\end{eqnarray}
In the above expression, given a unitary transformation $U$, we define ${\cal U}(\cdots)  = U (\cdots) U^\dag$, 
while we used the symbol 
 ${\cal M}^{\otimes n}$ to represent ${\cal M}_{E_1} \circ \cdots \circ {\cal M}_{E_n}$, ${\cal M}_{E_j}$ being the map ${\cal M}$ operating on the $j$-th element ${E}_j$ of ${\cal E}$. 
 The transformation  ${\cal R}^{(n)}_{S_m,{\cal E}}$ describes the evolution of $S_m$ in its interaction with ${\cal E}$ plus the subsequent free evolution of the latter induced by  the maps~$\cal M$.  
 Alternatively, exploiting the fact 
 that for $m'\neq m$, $n'\neq n$ the operators $U_{S_m, E_n}$ and $U_{S_{m'}, E_{n'}}$ commute, 
   ${\cal W}^{(n,M)}$ can also be expressed in terms of  {\em column} super-operators concatenated in series as follows:
\begin{eqnarray}\label{colu}
{\cal W}^{(n,M)}  =  {\cal C}^{(M)}_{{\cal S}, E_n} \circ {\cal C}^{(M)}_{{\cal S}, E_{n-1}} \circ \cdots \circ {\cal C}^{(M)}_{{\cal S}, E_2} \circ {\cal C}^{(M)}_{{\cal S}, E_1} \;,
 \end{eqnarray} 
 where for all $j =1, \cdots, n$,
 \begin{eqnarray}
 {\cal C}^{(M)}_{{\cal S}, E_j} &:=& {\cal M}_{ E_j} \circ {\cal U}_{S_{M},E_j}  \circ \cdots
   \circ {\cal M}_{ E_j}
 \circ {\cal U}_{S_{1},E_j}\;. \label{COLUMN}
 \end{eqnarray} 
 Equations~(\ref{ROW}) and (\ref{COLUMN}) enlighten the causal structure of the model. In particular (\ref{ROW}) makes it explicit that
whatever happens to  $S_{m+1}$ comes always {\em after} the transformations operating on $S_m$. As a consequence the latter can
have indirect influence on the former but the vice-versa is not allowed. Similarly Eq.~(\ref{COLUMN}) shows that an analogous causal structure
is present on the elements of ${\cal E}$: events involving $E_{n}$ may have causal influence on those involving $E_{n+1}$ but the opposite
is impossible. This last equation is also useful to write a recursive expression for $R(n)$. Indeed by construction we have 
\begin{eqnarray}
R(n+1)&=&{\cal C}^{(M)}_{{\cal S}, E_{n+1}}  (R(n)\otimes \eta)\;,\label{rec}
\end{eqnarray} 
which confirms the intrinsic Markovian structure of the temporal evolution for the {\em whole}  set of carriers that composes ${\cal S}$. 
The recursive form of Eq.~(\ref{rec}) is similar to the one characterizing the models 
of Refs.~\cite{ZSB}
 where, for a single qubit carrier ($M=1$) and for a  particular  class of interaction unitaries,
it was shown that  it  leads to a dynamics which can be described by a 
Lindblad super-operator. Following Ref.~\cite{PRL} one can generalize this fact to 
an arbitrary number of carriers and for arbitrary coupling Hamiltonians~(\ref{ggrss}).
We simply assume
 a weak coupling regime where we take a proper expansion with respect to the parameters $g$ and $\Delta t$ which quantifies the intensity and the duration of the single events. 
 In particular we  work in the regime in which  $g\Delta t$ is small enough to  allow for the expansion of the dynamical equation~(\ref{rec})
  up to  ${\cal O}\big((g\Delta t)^2\big)$, i.e. 
 \begin{eqnarray}
R(n+1)=
\big[ {\cal I}_{{\cal S}, E_{n+1}}&+&  {\cal C}^{\prime}_{{\cal S}, E_{n+1}} g \Delta t \\ \nonumber 
&+&  {\cal C}^{\prime\prime}_{{\cal S}, E_{n+1}} (g\Delta t)^2  \big]  (R(n)\otimes \eta)  + {\cal O}\big((g\Delta t)^3\big)
\;,\nonumber  \label{EQperR}
\end{eqnarray} 
where ${\cal I}_{{\cal S}, E_{n+1}}$ is the identity superoperator while ${\cal C}^{\prime}_{{\cal S}, E_{n+1}}$ and ${\cal C}^{\prime\prime}_{{\cal S}, E_{n+1}}$ are the first and second expansion terms in $g\Delta t$  of 
the superoperator  ${\cal C}^{(M)}_{{\cal S}, E_{n+1}}$, respectively (see below).
The resulting expression can then be traced over the degree of freedom of ${\cal E}$ to get an equivalent expression for the
the reduced density matrix of ${\cal S}$ alone,  yielding,  
  \begin{eqnarray}
\rho(n+1)&=&\rho(n) +  ( g\Delta t) \left\langle  {\cal C}^{\prime}_{{\cal S}, E_{n+1}}   \big(R(n)\otimes \eta\big) \right\rangle_{\cal E}   \label{EQperR111} \\
&& \qquad +  (g\Delta t)^2  \left\langle  {\cal C}^{\prime\prime}_{{\cal S}, E_{n+1}}  \big(R(n)\otimes \eta\big) \right\rangle_{\cal E} + {\cal O}\big((g\Delta t)^3\big)
\;,\nonumber 
\end{eqnarray} 
where we used the symbol  $\langle \cdots \rangle_{\cal E}$ to represent  the  partial trace over $E_1, E_2, \cdots$
and where for all $n$ we introduced 
\begin{eqnarray}
 \rho(n) :=\langle R(n)\rangle_{\cal E}\;, \end{eqnarray}
 (it represents the joint state of the carriers after the interaction with the first $n$ sub-environment).   
Explicit expressions  can be obtained by noticing that 
for each $m$ and $j$, the super-operators ${\cal U}_{S_{m},E_j}$ admit the following expansion,
\begin{eqnarray}
{\cal U}_{S_{m},E_j}  \label{compa}
= {\cal I}_{S_m,E_j} + (g\Delta t )\; {\cal U}_{S_{m},E_j}^\prime + (g\Delta t)^2 \; {\cal U}_{S_{m},E_j}^{\prime\prime} 
 + {\cal O}\big((g\Delta t)^3\big)\;, 
 \end{eqnarray}
 with 
 \begin{eqnarray} 
 {\cal U}_{S_{m},E_j}^\prime (\cdots) := 
 - i \Big[ H_{S_m,E_j} , (\cdots) \Big]_-   \;, \\
 {\cal U}_{S_{m},E_j}^{\prime\prime} (\cdots) := 
  H_{S_m,E_j} (\cdots ) H_{S_m,E_j}
 - \frac{1}{2} \Big[ H_{S_m,E_j}^2, (\cdots) \Big]_+ \;,  \label{compa1}
\end{eqnarray} 
where $[\cdots, \cdots]_-$ and $[\cdots, \cdots]_+$ represent the  commutator and the anti-commutator brackets respectively. 
From Eq.~(\ref{COLUMN}) it then follows that 
\begin{eqnarray}
 {\cal C}_{{\cal S}, E_j}^{\prime} &:=&
 \sum_{m=1}^M  {\cal M}_{ E_j}^{M-m+1} \circ {\cal U}_{S_{m},E_j}^\prime  
  \circ {\cal M}_{ E_j}^{m-1}\;, \label{COLUMN666} \\
   {\cal C}_{{\cal S}, E_j}^{\prime\prime} &:=&   {\cal C}_{{\cal S}, E_j}^{\prime\prime,a}   +    {\cal C}_{{\cal S}, E_j}^{\prime\prime,b}\;,
   \end{eqnarray}
   with 
   \begin{eqnarray}
 {\cal C}_{{\cal S}, E_j}^{\prime\prime,a}  &:= & \sum_{m=1}^M  {\cal M}_{ E_j}^{M-m+1} \circ {\cal U}_{S_{m},E_j}^{\prime\prime}   \circ {\cal M}_{ E_j}^{m-1}\;,   \nonumber \\
  {\cal C}_{{\cal S}, E_j}^{\prime\prime,b} &:= &
 \sum_{m' =m+1}^M  \sum_{m=1}^{M-1} {\cal M}_{ E_j}^{M-m'+1} \circ {\cal U}_{S_{m'},E_j}^{\prime}  
   \circ {\cal M}_{ E_j}^{m'-m}\circ {\cal U}_{S_{m},E_j}^{\prime}   \circ  {\cal M}_{ E_j}^{m-1}\;, \label{defCsec}
 \end{eqnarray} 
 (here ${\cal M}_{E}^{m}$ stands for the iterated application of $m$ maps ${\cal M}$ on the {\em same} sub-environmental system $E$, e.g.
 ${\cal M}_E^{2} = {\cal M}_E \circ {\cal M}_E$). 
Replacing  these expressions into Eq.~(\ref{EQperR111}) and remembering the definition~(\ref{ggrss})
 the first order term in $g\Delta t$ gives
\begin{eqnarray}
 \left\langle  {\cal C}^{\prime}_{{\cal S}, E_{n+1}}   \big(R(n)\otimes \eta\big) \right\rangle_{\cal E}
= -i \Big[ \sum_m H_{m}^{(eff)}, \rho(n)\Big]_-\;,
 \end{eqnarray} 
 with $H_{m}^{(eff)}$ being the following effective local Hamiltonians 
 \begin{eqnarray}\label{HEFFICACE}
 H_{m}^{(eff)}:=    \sum_{\ell}  \; \left\langle B^{(\ell)}_{E_{n+1}} {\cal M}_{ E_{n+1}}^{m-1} (\eta) \right\rangle_{E_{n+1}} \; A^{(\ell)}_{S_m}\;.
 \end{eqnarray}
 For the second order terms in $g\Delta t$ we get instead two contributions associated respectively to  
{\em local} Lindblad terms (i.e. Lindblad terms which act locally on the $m$-th carrier) and {\em two-body non local terms}  which couple the $m$ carrier to the $m' >m$.
More precisely  the first one is given by  
\begin{eqnarray}
 \left\langle  {\cal C}^{\prime\prime,a}_{{\cal S}, E_{n+1}}   \big(R(n)\otimes \eta\big) \right\rangle_{\cal E} 
= \frac{1}{\gamma} \; \sum_{m} {\cal L}_m (\rho(n)) \;,  \label{cprimeprimea} 
\end{eqnarray}
 where $\gamma$ is a positive parameter whose value will be specified later (see Eq.~(\ref{defgamma}) below), while 
 ${\cal L}_m$ is the super-operator 
 \begin{eqnarray}
 {\cal L}_m (\cdots) &= &\frac{1}{2} \sum_{\ell,\ell'} {\gamma_m^{(\ell, \ell')}} \big[ 2 A_{S_m} ^{(\ell')} (\cdots)A_{S_m} ^{(\ell)} 
 \nonumber \\
 &&-  A^{(\ell)}_{S_m}  A^{(\ell')}_{S_m}(\cdots)-  (\cdots)A^{(\ell)}_{S_m}  A^{(\ell')}_{S_m}   \big] \;.  \label{trieste2old1} \end{eqnarray}
In this expression  the coefficients 
  \begin{eqnarray}\label{trieste3old}
 \gamma^{(\ell,\ell')}_m := \gamma \; 
\langle B_E^{(\ell)} B_E^{(\ell')} \; {\cal M}_E^{m-1}(\eta)\rangle_E\;,
 \end{eqnarray}
 define the (non negative)  correlation matrix of the sub-environment operators $B_E^{(\ell)}$ and $B_E^{(\ell')}$
  evaluated  on the density matrix ${\cal M}^{m-1}(\eta)$ which describes the state of the sub-environment after $m-1$ free (i.e. non collisional) evolution steps.
  Equation~(\ref{trieste2old1}) can   also be casted in a more traditional form~\cite{PET}  by diagonalizing  $\gamma^{(\ell,\ell')}_m$:
  this allows one to identify the decay rates  of the system with the non-negative eigenvalues $r^{(\ell)}_{m}$ of $\gamma^{(\ell,\ell')}_m$ and the
associated Lindblad operators $L_{S_m}^{(\ell)}$  with a proper linear combinations of the  $A_{S_m}^{(\ell)}$.

The second contribution of order two in $g\Delta t$ which enters Eq.~(\ref{EQperR111}) is instead given by 
\begin{eqnarray}
 \left\langle  {\cal C}^{\prime\prime,b}_{{\cal S}, E_{n+1}}   \big(R(n)\otimes \eta\big) \right\rangle_{\cal E} = \frac{1}{\gamma} \;\sum_{m'>m} \; {\cal D}^{(\rightarrow)}_{m,m'} (\rho(n)) \;,  \label{cprimeprimeb}
 \end{eqnarray} 
  where for $m'>m$ 
   ${\cal D}^{(\rightarrow)}_{m,m'}$ is the super-operator  defined as
 \begin{eqnarray}
{\cal D}^{(\rightarrow)}_{m,m'} (\cdots) &=& \sum_{\ell, \ell'} \gamma_{m,m'}^{(\ell, \ell')}  \;\;  A_{S_m}^{(\ell)}  \; \Big[ (\cdots) ,A_{S_{m'}}^{(\ell')} \Big]_- 
\nonumber \\
&&- \sum_{\ell, \ell'}   [\gamma_{m,m'}^{(\ell, \ell')}]^* \;  \;   \Big[(\cdots), A_{S_{m'}}^{(\ell')}  \Big]_- \;A_{S_m}^{(\ell)}\;,
\label{d12}
\end{eqnarray}
with 
   \begin{eqnarray} \label{gamma}
    \gamma^{(\ell,\ell')}_{m,m'}  :=   \gamma\; 
\langle B_{E}^{(\ell')} {\cal M}^{m'-m}( B_{E}^{(\ell)} \; {\cal M}^{m-1}(\eta)) \; \rangle_{E}\;.
 \end{eqnarray}
The coefficients $\gamma^{(\ell,\ell')}_{m,m'}$ introduce
 cross correlations among the carriers and depend  upon the {\em distance} $m'-m$ between the associated rows of the graph~of Fig.~\ref{fig1}. 
Furthermore, similarly to the the terms of Eq.~(\ref{trieste3old}), they  also depend on $m-1$ due to the fact that the model admits a {\em first} carrier.

 The resulting expression for $\rho(n+1)$ can thus be written as 
 \begin{eqnarray}
&& \frac{\rho(n+1)-  \rho(n)}{\Delta t}  =  -i g \Big[ \sum_m H_{m}^{(eff)}, \rho(n)\Big]_-  \nonumber \\
 &&\qquad +
  \frac{g^2 \Delta t }{\gamma} \; \left\{  \sum_{m}  {\cal L}_m (\rho(n) )     
 + \sum_{m'>m} {\cal D}^{(\rightarrow)}_{m,m'} (\rho(n) )  \right\} 
\label{EQperR11245}   \;  + {\cal O}\big(g^3\Delta t^2\big) \;,
\end{eqnarray} 
which possesses an explicit Markovian structure characterized by the presence of an effective Hamiltonian (first line) and dissipative contributions (second line).
\newline

Before proceeding to the continuos limit let us briefly review
how the above scheme applies to a specific discrete system, with the aim to clarify the meaning and the validity of the assumptions made in our model. For this purpose we 
refer to the prototypical example of Fig.~\ref{fig3}~c).
 Here  an array of cavities is driven by a beam of atoms crossing them. The rate of injection of the atoms is such that the atoms enter the cavity one by one as shown in 
 figure. The atoms in the beam, all initially prepared in the same state  cross sequentially all the cavities of the array. The atom-field interaction is described by the Jaynes-Cummings hamiltonian, which can be straightforwardly cast in the form (\ref{ggrss}). For a single cavity crossed by a beam of single atoms, in the absence of leackage of photons out of the cavity, such model describes damping or amplification for the cavity field, whose dynamics can be described by a Markovian master equation in Lindblad form $\cite{Schleich}$.  The extension to  $n$ cavity is described naturally in our model. 
Such markovian  behavior  is due to two crucial features of the model: the short (finite) time $\Delta t$ which takes each atom to cross a cavity and the fact that the atomic state is ``refreshed" when a new atom is injected. The master equation so obtained  describes a coarse grained time derivative on a timescale $\Delta t$. On such time scale the environment is reset. In the standard theory of damping of a system which is continuosly interacting with the same - big - reservoir the time $\Delta t$ would be the self correlation time of the reservoir. This is not the case in our scenario: each sub environment is small but it interacts for a short time and then, after a time $\Delta t$ substituted with a new one. Furthermore the cross terms in our master equation do not describe a collective, simultaneous, coupling of the subsystems with the environment, but rather they describe how the dynamics of the various subsystems are correlated due to the fact that they have interacted sequentially with the sub environments. This explains why the dynamics described by our system is markovian and does not exhibit the non markovian multipartite features which are characteristic of the scenarios analyzed in \cite{Gordon}.

\subsection{Continuous limit}\label{sec:continuous}
Equation~(\ref{EQperR11245}) can be turned into a continuos time expression by taking the proper limit $\Delta t \rightarrow 0$ while sending  $n$ to infinity so that 
\begin{eqnarray} \label{limitcont}
\lim_{\Delta t  \rightarrow 0^+} n \; \Delta t = t \;.
\end{eqnarray}
Notice that there are two possible regimes. If $g$ is kept constant as $\Delta t$ goes to zero, then the dissipative contributions of Eq.~(\ref{EQperR11245}) are washed away and the dynamics reduces to a unitary
 evolution characterized by the effective (possibly time-dependent) Hamiltonian~(\ref{HEFFICACE}). The situation becomes more
 interesting if  instead $g$ is sent to infinity so that 
$g^2 \Delta t$ remains finite,  i.e.~\cite{PRL}
\begin{eqnarray} 
 \lim_{\Delta t  \rightarrow 0^+} g^2 \Delta t = \gamma \;.
\label{defgamma} 
\end{eqnarray} 
Enforcing this limit is of course problematic due to the presence of the first order contribution in Eq.~(\ref{EQperR11245})  which
tends to explode. A way out  is to assume the following {\em stability} condition for the environment~\cite{PRL}, 
\begin{eqnarray}~\label{ASSUMPTION1}
\langle B_E^{(\ell)} {\cal M}^{m} (\eta) \rangle_E = 0\qquad  \forall \ell, m \;,
\end{eqnarray} 
which ensures that $H_{m}^{(eff)}$, and hence the first order contribution of Eq.~(\ref{EQperR11245}), identically nullifies. 
Under this hypothesis,  defining $\rho(t)= \lim_{\Delta t \rightarrow 0^+} \;  \rho(n)$,
 one can indeed arrive to the following continuous  master equation for the system, 
\begin{eqnarray}
{\dot \rho}(t) = 
 \sum_{m=1} {\cal L}_m (\rho(t) )     
+ \sum_{m'>m} {\cal D}^{(\rightarrow)}_{m,m'} (\rho(t) ) \;, \label{fin}
 \end{eqnarray}  whose properties have been characterized in Ref.~\cite{PRL}. Here we only mention 
  that the cross terms appearing in Eq.~(\ref{fin}) 
  have an intrinsic unidirectional character which makes this expression  
 suitable to characterize the
  dynamics of a cascade quantum system. Indeed, for each 
 $m'>m$ it can be directly verified from Eq.~(\ref{d12}) that we have
 \begin{eqnarray} \label{prop1}
\left\langle {\cal D}^{(\rightarrow)}_{m,m'} (\cdots)\right\rangle_{S_{m'}} = 0 \;.
\end{eqnarray} 
This  implies that the evolution of the first $m$ carriers of the system are not 
influenced by the evolution of the ones that follow (in other words it is possible to write a master equation for the density matrix of the
 first $m$ elements of ${\cal S}$ only). 
The opposite relation however is not true  as in general $ {\cal D}^{(\rightarrow)}_{m,m'}$  doesn't nullify  
when traced over $S_{m}$, i.e. 
$\left\langle {\cal D}^{(\rightarrow)}_{m,m'} (\cdots)\right\rangle_{S_{m}}\neq 0$. This means in particular that in our model
it is in general  impossible to write a master equation that involves {\em only} the density matrix of the $m'$-th carrier $S_{m'}$ (we need indeed to include also the
carriers that precede it\footnote{A notable exceptions being the case in which  all the coefficients $\gamma^{(\ell,\ell')}_{m,m'}$ appearing in Eq.~(\ref{d12})
are real: under this condition $\left\langle {\cal D}^{(\rightarrow)}_{m,m'} (\cdots)\right\rangle_{S_{m}}=0$ so that 
the dynamics of every carrier is causally disconnected from the others~\cite{PRL}.}).  
Similar properties were obtained in the works of Gardiner, Parkins and Carmichael~\cite{G} in theirs seminal study of 
cascade optical quantum systems. As shown in Ref.~\cite{PRL} the latter can be seen as special instances of (\ref{fin}) for specific 
choices of the couplings~(\ref{ggrss}) and of the environment initial state $\eta$.

In what follows we will not discuss  further the implications of Eq.~(\ref{prop1}). Instead we will focus on the 
assumption Eq.~(\ref{ASSUMPTION1}) showing how it can be enforced by passing in a proper interaction picture with respect to the free evolution
of the carriers. Before doing so however we think it is worth stressing 
  that the above derivation 
still holds also if the collisional Hamiltonians (\ref{ggrss}) are not uniform. For instance suppose we have
\begin{eqnarray}\label{hamnonuni}
H_{S_mE_n} := \sum_\ell \; A^{(n,\ell)}_{S_m}  \otimes B^{(m,\ell)}_{E_n} \;,
\end{eqnarray} 
where now the operators acting on the carrier $S_m$ are allowed to explicitly depends upon 
the $n$ index which label the collisional events, and similarly the operators acting on the sub-enviroment are allowed
to explicitly depends upon the index $m$ which labels the carriers.
 Under these conditions one can verify that  Eq.~(\ref{EQperR11245}) is still valid even though 
 both  the  super-operators  ${\cal L}_m$ and ${\cal D}^{(\rightarrow)}_{m,m'}$ become explicit functions of the carriers labels {\em and} of the
index $n$ which  plays the role of a temporal parameter for the reduced density matrix $\rho(n)$. Specifically they are now 
defined respectively as in Eqs.~(\ref{cprimeprimea}) and (\ref{cprimeprimeb})
with the operators  $A^{(n+1,\ell)}_{S_m}$ instead of $A^{(\ell)}_{S_m}$ and with
the coefficients $\langle B_E^{(\ell)} B_E^{(\ell')} \; {\cal M}^{m-1}(\eta)\rangle_E$ and 
$ \langle B_{E}^{(\ell')} {\cal M}^{m'-m}( B_{E}^{(\ell)} \; {\cal M}^{m-1}(\eta)) \rangle_{E}$ replaced by
$\langle B_E^{(m,\ell)} B_E^{(m,\ell')} \; {\cal M}^{m-1}(\eta)\rangle_E$ and 
$ \langle B_{E}^{(m',\ell')} {\cal M}^{m'-m}( B_{E}^{(m,\ell)} \; {\cal M}^{m-1}(\eta))  \rangle_{E}$ 
respectively.
Similarly the continuous limit can be enforced as before: in this case however, to account for the non uniformity of the couplings, 
the condition~(\ref{ASSUMPTION1}) becomes
\begin{eqnarray}\label{ASSUMPTION2}
\left\langle B^{(m,\ell)}_{E} {\cal M}_{ E}^{m-1} (\eta) \right\rangle_{E} =0, \quad  \forall m, \ell \;.
\end{eqnarray} 
Furthermore, while taking the limit (\ref{limitcont})
the operators $A^{(n+1,\ell)}_{S_m}$  acquire an explicit temporal dependence 
 which   transforms them into  a 
one parameter family of operators.
As a result we get a 
 time-dependent  master equation characterized by a Lindblad generators which explicitly depends
on~$t$, i.e. 
\begin{eqnarray}
{\dot \rho}(t) = 
 \sum_{m=1} {\cal L}_m (\rho(t) ; t )     
 + \sum_{m'>m} {\cal D}^{(\rightarrow)}_{m,m'} (\rho(t);t ) \;, \label{fintimedep}
 \end{eqnarray}
with ${\cal L}_m(\cdots; t)$ and ${\cal D}^{(\rightarrow)}_{m,m'} (\cdots;t )$ 
as in Eqs.~(\ref{trieste2old1}) and (\ref{d12}) with the operators $A^{(\ell)}_{S_m}$ replaced by 
$A^{(\ell)}_{S_m}(t):=\lim_{\Delta t\rightarrow 0^+} A^{(n+1,\ell)}_{S_m}$.

\section{The stability condition}\label{stabilitycond}
The name {\em stability} condition given to  the constraint~(\ref{ASSUMPTION1})
follows from the fact that it implicitly assumes that during the collisions the sub-environments  are not affected  by the coupling with the carriers (at least  at first order in the coupling strength). 
This is mathematically equivalent to the standard derivation of a Markovian master equation~\cite{PET}
for a system interacting with a large environment, in which one assumes that  the overall system-environment density operator 
at any given time $t$ of the evolution factorizes as in $\rho(t) \otimes \eta$ where  $\eta$ is the environment density operator. The two scenarios are however different. In the standard case the reason for which the environment state is unchanged is because it is big. In the scenario analyzed here, consistently with the collisional model, the environment state is constant because each subsystem collides briefly with a sequence of sub-environments all initially in the same state. 

As anticipated in the previous section, in our analysis of the condition~(\ref{ASSUMPTION1}) a proper handling of the carriers free evolution plays a fundamental role. This should not
come as a surprise: an important  step in the standard derivation of a Lindblad form is the possibility of 
effectively ``removing"  the free evolution of the system and of the environment
by passing in the associated interaction representations. Such step is useful because it allows one to directly relate the fast evolution times of the large environment with 
the slow decaying rates of the system of interests: it is in this limit that the Markov approximation can be properly enforced~\footnote{
The need of removing  the free evolution of ${\cal S}$ from the description of the system dynamics is clearly evident also in our case. 
Indeed the condition (\ref{ASSUMPTION1}) is clearly incompatible with the presence of free local contributions in the Hamiltonians $H_{S, E}$ as they will 
correspond to terms of the form $H_{S}^{(free)}\otimes I_{E}$, i.e.  contributions $ A^{(\ell)}_{S}  \otimes B^{(\ell)}_{E}$ with $B_{E}^{(\ell)}$ being the
identity operator which will yield $\langle B_E^{(\ell)} {\cal M}^{m} (\eta) \rangle_E =\langle  {\cal M}^{m} (\eta) \rangle_E 
 = 1$.}.  In our model we can show that the cases in which Eq.~(\ref{ASSUMPTION1}) cannot be directly enforced, can be mapped into 
effective models in which Eq.~(\ref{ASSUMPTION1}) exactly holds but which allows for explicit free evolution terms for the carriers 
between any two collisions which have to be removed by passing in a proper interaction picture representation. 
  As a preliminary step toward the discussion of the stability condition  it is hence important to discuss how the derivation changes in this last
  circumstance.

\subsection{Including local free evolution terms for the carriers}\label{sec:includ}

Assume that the stability condition~(\ref{ASSUMPTION1}) holds, but that 
 between any two consecutive collisions, the carriers undergo to a free-evolution described
by a (possibly time-depedent)  Hamiltonian $H_{{\cal S}}(t): =\sum_{m} h_{S_m}(t)$ which are local (i.e. no direct interactions between the carriers is allowed).
Under this circumstance  it is possible to show that 
 Eq.~(\ref{fin}) still holds in the proper interaction picture representation at the price of
allowing the generators of the resulting master equation to be explicitly time dependent as in Eq.~(\ref{fintimedep}).

To see this we first notice that under the assumption that the collision time $\Delta t$  is much shorter than the time interval that elapses between two consecutive
collisional events  (i.e. $\Delta t \ll \tau_{n}-\tau_{n-1}$),
the unitary operator which describes the evolution of the $m$-th carrier in its interaction with ${\cal E}$  is now given by 
\begin{eqnarray}\label{hh2222}
&&U_{S_m {\cal E}}^{(n)}
 :=  U_{S_m,E_n} V_{S_m}( \tau_n, \tau_{n-1})\;U_{S_m,E_{n-1}} \\ \nonumber
 &&\qquad \qquad \;\cdots \; V_{S_m}(\tau_2,\tau_1)\;U_{S_m,E_2} \; V_{S_m}(\tau_1,0) \;U_{S_m,E_1},
\end{eqnarray} 
where $U_{S_m,E_n}$ are the collisional transformations, $\tau_n$ is the time at which the $n$-th collision takes place, and where 
$V_{S_m}(\tau_{n},\tau_{n-1}):={\cal T}
 \exp[-i \int_{\tau_{n-1}}^{\tau_n} dt' h_{S_m}(t')]$ is the 
unitary operator which describes the free-evolution of $S_m$ between the $(n-1)$-th and the $n$-th collision
(in this expression ${\cal T}
 \exp[\cdots]$ indicates the time-ordered exponential which we insert to explicitly account for possibility that  the $h_{S_m}$ will be 
  time-dependent). 
Define hence the operators 
\begin{eqnarray}
\bar{A}^{(n,\ell)}_{S_m}  : =  V_{S_m}^\dag (\tau_n,0)  \;  {A}^{(\ell)}_{S_m}\;   V_{S_n} (\tau_n,0)\;,
\end{eqnarray} and
the
Hamiltonian 
\begin{eqnarray} \label{fffs}
\bar{H}_{S_m,E_n} := 
 V_{S_n}^\dag (\tau_n,0) \; {H}_{S_m,E_n} V_{S_n}  (\tau_n,0)= \sum_\ell \; \bar{A}^{(n,\ell)}_{S_m} \otimes B^{(\ell)}_{E_n} 
\;,
\end{eqnarray}
which describes  the coupling between $S_m$ and ${\cal E}$ in the interaction representation associated with the free evolution of $S_m$.
Notice that the  operators $\bar{A}^{(n,\ell)}_{S_m}$ are explicit  functions of the index $n$ which labels the collisions as in the case of Eq.~(\ref{hamnonuni}) (here
however the terms operating on ${\cal E}$ are kept uniform). 
Observing that  for all $\ell$  one has $V_{S_m}(\tau_\ell,\tau_{\ell-1}) V_{S_m}(\tau_{\ell-1},\tau_{\ell-2}) = V_{S_m}(\tau_\ell,\tau_{\ell-2})$
we  can now write Eq.~(\ref{hh2222}) as
\begin{eqnarray}\label{hh22223}
&&U_{S_m {\cal E}}^{(n)}
 := V_{S_m}(\tau_{n},0) \;  \bar{U}_{S_m,{\cal E}}^{(n)},
\end{eqnarray} 
where $\bar{U}_{S_m,{\cal E}}^{(n)}$ is the unitary that defines the collisions of $S_m$  with the sub-environments in the interaction representation, i.e.  
\begin{eqnarray}\label{hh2222555}
&&\bar{U}_{S_m,{\cal E}}^{(n)}
 :=  \bar{U}_{S_m,E_n} \;\bar{U}_{S_m,E_{n-1}}\cdots  \bar{U}_{S_m,E_1},
\end{eqnarray} 
with 
\begin{eqnarray} 
 \bar{U}_{S_m,E_n}  = \exp[ - i g\;  \bar{H}_{S_m,E_n} \Delta t ]\;.
 \end{eqnarray} 
 Similarly we can express   the super-operators ${\cal W}^{(n,M)}$  as 
\begin{eqnarray}
{\cal W}^{(n,M)}  &=&   {\cal V}_{{\cal S}}(\tau_{n},0) \circ \; \bar{\cal W}^{(n,M)} \;, \\
\bar{\cal W}^{(n,M)} &:=&   \bar{\cal C}^{(M)}_{{\cal S}, E_n}  \circ \cdots  \circ \bar{\cal C}^{(M)}_{{\cal S}, E_1}\;, \\
 \bar{\cal C}^{(M)}_{{\cal S}, E_j} &:=& {\cal M}_{ E_j} \circ \bar{\cal U}_{S_{M},E_j} \circ \cdots
   \circ {\cal M}_{ E_j}
 \circ \bar{\cal U}_{S_{1},E_j},
\end{eqnarray}
with ${\cal V}_{{\cal S}}(\tau_{n},0)$ being the super-operator associated with the
joint free unitary evolution   obtained by combining all the local terms  of the carriers, i.e. 
$V_{\cal S}(\tau_{n},0) := V_{S_1}(\tau_{n},0) \cdots V_{S_M}(\tau_{n},0)$.
 Defining hence $\bar{R}(n)$ the state of ${\cal S}$ and of the first elements of ${\cal E}$ after $n$ collisions in the interaction representation induced
by $V_{\cal S}(\tau_{n},0)$ as 
\begin{eqnarray} 
\bar{R}(n) =V_{\cal S}^\dag(\tau_n,0) \; R(n) \; V_{\cal S}(\tau_n,0)\;,
\end{eqnarray} 
we get a recursive expression analogous to Eq.~(\ref{rec}) with ${\cal C}^{(M)}_{{\cal S}, E_{n+1}}$ replaced by 
$\bar{\cal C}^{(M)}_{{\cal S}, E_{n+1}}$, i.e. 
 \begin{eqnarray}
\bar{R}(n+1)&=&
\bar{\cal C}^{(M)}_{{\cal S}, E_{n+1}}  (\bar{R}(n)\otimes \eta)\;. \label{rec1}
\end{eqnarray} 
More precisely this expression coincides with that which, as in the case described at the end of Sec.~\ref{sec:continuous}, one would have obtained starting from a collisional model
in which no free evolution of the carriers is allowed but the collisional events are not uniform. Indeed the generators of the dynamics $\bar{H}_{S_m,E_n}$ do have 
the same form of the Hamiltonians~(\ref{hamnonuni}). Following the same prescription given there, 
we can then get an expression for the reduced density matrix 
 $\bar{\rho}(n)=\langle \bar{R}(n) \rangle_{\cal E}$  which represents the state of the carriers after $n$ collisions in the interaction picture with respect to the free evolution 
 generated by $H_{\cal S}(t)$.
Enforcing the limit~(\ref{defgamma}) under the condition~(\ref{ASSUMPTION2}),  one can verify that  $\bar{\rho}(t)$ obeys 
to a ME analogous to Eq.~(\ref{fin}) with the operators $A_{S_m} ^{(\ell')}$ being replaced by 
the time-dependent operators $\bar{A}^{(\ell)}_{S_m}(t):= \lim_{\Delta t\rightarrow 0^+} \bar{A}^{(n,\ell)}_{S_m}$.

\subsection{Enforcing the stability condition via a global unitary mapping}\label{sec:enforcing}
 Now that we have learned how to deal with free local evolutions terms operating between the collisional events, we
 show how to use this result to effectively enforce the stability condition of Eq.~(\ref{ASSUMPTION1}) for models in which it doesn't apply rigorously. 
Specifically we shall see that  such condition can be imposed  by first moving into an interaction representation with respect to a rescaled
local Hamiltonian for the system ${\cal S}$ which maps the problem into one equivalent to that discussed in Sec.~\ref{sec:includ}. 

Indeed let 
\begin{eqnarray}\label{ggrss1}
 H_{S_mE_n} := \sum_\ell \; A^{(n,\ell)}_{S_m} \otimes B^{(\ell)}_{E_n} \;,
 \end{eqnarray} 
be the Hamiltonian which describe the collisions between the carriers and the sub-environments (notice that 
we are allowing the operators $A^{(n,\ell)}_{S_m}$ to depend explicitly from the $n$ label to account for possible  
local free evolution of the carriers as discussed in the previous section). 
Suppose then that Eq.~(\ref{ASSUMPTION1}) does not hold. In this case we define 
\begin{eqnarray}
B^{(m,\ell)}_{E}  &:=&  B^{(\ell)}_{E} - \delta_m^{(\ell)}  \; I_{E}\;,\\
\delta_m^{(\ell)}  &:=& \langle B_E^{(\ell)} {\cal M}^{m-1} (\eta) \rangle_E\;,
\end{eqnarray} 
and write,
\begin{eqnarray}\label{ggrss1000}
 H_{S_m, E_n}&:=& \Delta H_{S_m,E_n} + h_{S_m}^{(n)}\;,
 \end{eqnarray}
 where 
 \begin{eqnarray}
 h_{S_m}^{(n)} := \sum_\ell \delta_m^{(\ell)} \; A^{(n,\ell)}_{S_m} \otimes   I_{E_n}\;,
 \end{eqnarray}
  is a local Hamiltonian on $S_m$ while 
\begin{eqnarray}
\Delta H_{S_m,E_n} :=  \sum_\ell \; A^{(n,\ell)}_{S_m}  \otimes B^{(m,\ell)}_{E_n}\;,
\end{eqnarray} 
is a rescaled coupling Hamiltonian. Differently from the original one given in Eq.~(\ref{ggrss1}), but similarly to Eq.~(\ref{hamnonuni}), it is built from operators $B^{(m,\ell)}_{E_n}$ which
explicitly depend on the label $m$ of the carrier $S_m$, and which {\em by construction} satisfy the generalized condition~(\ref{ASSUMPTION2}), i.e. 
\begin{eqnarray} \label{IMPO}
 \langle B_E^{(m,\ell)} \; {\cal M}^{m-1} (\eta) \rangle_E =0 \;.
 \end{eqnarray} 
 Passing then in the  interaction representation with respect to $h_{S_m}^{(n)}$ we can thus express the unitary evolution induced by $H_{S_m, E_n}$ as
\begin{eqnarray}
U_{S_m, E_n} =\exp[ -i g H_{S_m,E_n}  \Delta t]  =
 e^{-i g  h_{S_m}^{(n)} \Delta t }
 \;  {\cal T} \exp[ -i g \int_0^{\Delta t} dt'  \; \overline{\Delta H}_{S_m,E_n}(t') ] \;, \nonumber 
\end{eqnarray}  
where $e^{-i g  h_{S_m}^{(n)} \Delta t }$ is a local unitary on $S_m$ while 
\begin{eqnarray}
 \overline{\Delta H}_{S_m,E_n}(t) &:=&   e^{i g  h_{S_m}^{(n)} t } \;  {\Delta H}_{S_m,E_n} \;  e^{-i g  h_{S_m}^{(n)} t }\;. \nonumber 
  \end{eqnarray} 
 Therefore the rhs of Eq.~(\ref{hh2}) can be written now as 
 \begin{eqnarray}
U_{S_m, {\cal E}}^{(n)}= V_{S_m}^{(n)}  \;\tilde{U}_{S_m,E_{n}} \;\cdots \; \tilde{U}_{S_m,E_2}\; \tilde{U}_{S_m,E_1}\;,
\end{eqnarray} 
 where $V_{S_m}^{(n)}$ 
 and $\tilde{U}_{S_m,E_j}$ are the following unitary operators
 \begin{eqnarray}
 V_{S_m}^{(n)} 
 := e^{-i g  h_{S_m}^{(n)} \Delta t } \;e^{-i g  h_{S_m}^{(n-1)} \Delta t }
  \cdots  e^{-i g  h_{S_m}^{(2)} \Delta t }\;  e^{-i g  h_{S_m}^{(1)} \Delta t }\;, 
 \end{eqnarray}
 and
 \begin{eqnarray}\label{pp} 
  \tilde{U}_{S_m,E_j}:= 
    [ V_{S_m}^{(j-1)}]^\dag \;  \left(  {\cal T} \exp[ -i g \int_0^{\Delta t} dt'  \; \overline{\Delta H}_{S_m,E_j}(t') ] \right) \; V_{S_m}^{(j-1)}.
   \end{eqnarray}
   For future reference it is worth anticipating that  the term within the round brackets
    admits the following expansion in $\Delta t$,
   \begin{eqnarray} \nonumber 
 \fl \qquad 
 \quad I_{S_m,E_n} - i (g \Delta t)\; 
  {\Delta H}_{S_m,E_j} -  \frac{1}{2} (g \Delta t)^2  \left({\Delta H}_{S_m,E_j}  \right)^2 - \frac{i}{2}  (g \Delta t)^2  \;  {Q}_{S_m,E_j}  
     + {\cal O}(\Delta t^3) \;,
   \end{eqnarray}
 where the last contribution originates  from  the time-ordering in the exponential of Eq.~(\ref{pp}) and it is defined in terms of 
 the first derivative of ${\overline{\Delta H}}_{S_m,E_j}(t)$, i.e. 
 \begin{eqnarray}
 {Q}_{S_m,E_j}  :=  \frac{\partial \; {\overline{\Delta H}}_{S_m,E_j}(t)}{\partial t} \Big|_{t=0} = i g \Big[ h_{S_m}^{(j)},  {{\Delta H}}_{S_m,E_j} \Big]_-.
 \end{eqnarray} 
  This yields the  following expansion for the super-operator $\tilde{\cal U}_{S_{m},E_j}$ associated to the unitary ${\tilde U}_{S_{m},E_j}$, 
  \begin{eqnarray}
\tilde{\cal U}_{S_{m},E_j}  
= {\cal I}_{S_m,E_j} + (g\Delta t )\; \tilde{\cal U}_{S_{m},E_j}^\prime + (g\Delta t)^2 \; \tilde{\cal U}_{S_{m},E_j}^{\prime\prime} 
 + {\cal O}\big((g\Delta t)^3\big)\;,  \label{expnew}
 \end{eqnarray}
 with ${\cal I}_{S_m,E_j}$ being the identity map and with 
 \begin{eqnarray} 
 \fl \; \tilde{\cal U}_{S_{m},E_j}^\prime (\cdots)&:=& - i \Big[ 
 \Delta \hat{H}_{S_m,E_j} , (\cdots) \Big]_-   \;, \label{NEWREVE1} \\
\fl  \;  \tilde{\cal U}_{S_{m},E_j}^{\prime\prime} (\cdots)&:=&  \Delta \hat{H}_{S_m,E_j} (\cdots ) \Delta \hat{H}_{S_m,E_j} 
 - \frac{1}{2} \Big[ \Delta \hat{H}_{S_m,E_j}^2, (\cdots) \Big]_+ - \frac{i}{2} \Big[\hat{Q}_{S_m,E_j}  , (\cdots) \Big]_- \;,  \label{NEWREVE2} 
\end{eqnarray} 
where for easy of notation in this expression given a generic operator $\Theta_{S_m,E_j}$ we used the notation $\hat{\Theta}_{S_m,E_j}$  to represent 
its evolution via the unitary $V_{S_m}^{(j-1)}$, i.e. 
\begin{eqnarray}
\hat{\Theta}_{S_m,E_j}:= 
[ V_{S_m}^{(j-1)}]^\dag\Theta_{S_m,E_j} V_{S_m}^{(j-1)}\;.
\end{eqnarray}
The expression Eq.~(\ref{NEWREVE2}) should be compared with Eq.~(\ref{compa1}): we notice that due to the presence of the time-ordering in Eq.~(\ref{pp})
an extra term is present in the decomposition. We shall see however that when tracing out the sub-environments, such term plays no role in the evolution of the carries (see Eq.~(\ref{NEWIDENTITY}) below).

With the above  identities the row super-operator  ${\cal R}^{(n)}_{S_m,{\cal E}}$ entering in Eq.~(\ref{ROW}) can thus be expressed as 
 \begin{eqnarray}
{\cal R}_{S_m,{\cal E}} := {\cal M}^{\otimes n} \circ \left({\cal V}_{S_m}^{(n)} \circ  \;\tilde{\cal U}_{S_m,E_{n}}\circ \cdots \circ  \tilde{\cal U}_{S_m,E_1}\right) 
= {\cal V}_{S_m}^{(n)} \circ \tilde{\cal R}_{S_m,{\cal E}}\;,
\end{eqnarray} 
where as usual ${\cal V}_{S_m}^{(n)}$ and $\tilde{\cal U}_{S_m,E_{j}}$ represent the super-operators associated with the unitary transformations  ${V}_{S_m}^{(n)}$ and $\tilde{U}_{S_m,E_{j}}$
respectively, and where 
 \begin{eqnarray}
\tilde{\cal R}_{S_m,{\cal E}} &:=& {\cal M}^{\otimes n} \circ  \;\tilde{\cal U}_{S_m,E_{n}} \circ \cdots \circ  \tilde{\cal U}_{S_m,E_1}\;.\end{eqnarray} 
Accordingly Eq.~(\ref{ROW}) becomes
\begin{eqnarray}\label{ROW1}
{\cal W}^{(n,M)}  =   {\cal V}_{\cal S}^{(n)} \circ \tilde{\cal W}^{(n,M)}\;,
\end{eqnarray}
with  ${\cal V}_{\cal S}^{(n)}$ being the super-operator associated with the joint unitary 
${V}_{\cal S}^{(n)}:= {V}_{S_M}^{(n)}\otimes \cdots \otimes {V}_{S_1}^{(n)}$  and with
\begin{eqnarray}\label{ROW33}
 \tilde{\cal W}^{(n,M)}:= 
 \tilde{\cal R}_{S_M, {\cal E}} \circ \tilde{\cal R}_{S_{M-1}, {\cal E}} \circ  \cdots\circ   \tilde{\cal R}_{S_2, {\cal E}} \circ  \tilde{\cal R}_{S_1, {\cal E}} \;.
\end{eqnarray}
This can also be written in terms of column super-operators as in Eq.~(\ref{colu}). In particular we get 
 \begin{eqnarray}\label{colu1}
\tilde{\cal W}^{(n,M)}  =  \tilde{\cal C}_{{\cal S}, E_n} \circ \tilde{\cal C}_{{\cal S}, E_{n-1}} \circ \cdots \circ \tilde{\cal C}_{{\cal S}, E_2} \circ \tilde{\cal C}_{{\cal S}, E_1} \;,
 \end{eqnarray} 
with 
 \begin{eqnarray}
 {\cal C}_{{\cal S}, E_j} &:=& {\cal M}_{ E_j} \circ \tilde{\cal U}_{S_{M},E_j}  \circ \cdots
  \circ {\cal M}_{ E_j}
 \circ  \tilde{\cal U}_{S_{1},E_j}\;. \label{COLUMN11}
 \end{eqnarray} 
Defining now
\begin{eqnarray}\label{DEFDEF}
\bar{R}(n) := {\cal V}_{\cal S}^{(n)} (R(n)) = [{V}_{\cal S}^{(n)}]^\dag R(n) {V}_{\cal S}^{(n)} \;, 
\end{eqnarray} 
the state of the carriers and of the first $n$ sub-environemental state in the interaction
picture representation induced by the unitary ${V}_{\cal S}^{(n)}$, we have
\begin{eqnarray}
{\bar{R}}(n+1) = \tilde{\cal W}^{(n+1,M)}  (\rho(0) \otimes\eta^{\otimes {n+1}})= \tilde{\cal C}_{{\cal S}, E_{n+1}}
({\bar{R}}(n)\otimes \eta)  \;.
\end{eqnarray}
Take then the partial trace over ${\cal E}$ of this expression and use Eq.~(\ref{expnew}) to expand $\tilde{\cal C}_{{\cal S}, E_{n+1}}$.
Defining ${\bar{\rho}}(n)= \langle {\bar{R}}(n)\rangle_{\cal E}$ we get 
  \begin{eqnarray}\label{exp1}
&&{\bar{\rho}}(n+1)={\bar{\rho}}(n) +  ( g\Delta t) \left\langle  \tilde{\cal C}^{\prime}_{{\cal S}, E_{n+1}}   \big({\bar{R}}(n)\otimes \eta\big) \right\rangle_{\cal E}   \\
&& \qquad +  (g\Delta t)^2  \left\langle  \tilde{\cal C}^{\prime\prime}_{{\cal S}, E_{n+1}}  \big({\bar{R}}(n)\otimes \eta\big) \right\rangle_{\cal E} + {\cal O}\big((g\Delta t)^3\big)
\;,\nonumber 
\end{eqnarray} 
where $\tilde{\cal C}^{\prime}_{{\cal S}, E_{n+1}}$ and $\tilde{\cal C}^{\prime\prime}_{{\cal S}, E_{n+1}}$ are respectively the first and second order term of the
expansion of 
$\tilde{\cal C}_{{\cal S}, E_{n+1}}$, i.e. 
\begin{eqnarray}
 \tilde{\cal C}_{{\cal S}, E_j}^{\prime} &:=&
 \sum_{m=1}^M  {\cal M}_{ E_j}^{M-m+1} \circ \tilde{\cal U}_{S_{m},E_j}^\prime  
  \circ {\cal M}_{ E_j}^{m-1}\;, \label{COLUMN777} \\
  \tilde{\cal C}_{{\cal S}, E_j}^{\prime\prime} &:=&   \tilde{\cal C}_{{\cal S}, E_j}^{\prime\prime,a}   +    \tilde{\cal C}_{{\cal S}, E_j}^{\prime\prime,b}\;,
   \end{eqnarray}
   with 
   \begin{eqnarray}
 \tilde{\cal C}_{{\cal S}, E_j}^{\prime\prime,a} & := & \sum_{m=1}^M  {\cal M}_{ E_j}^{M-m+1} \circ \tilde{\cal U}_{S_{m},E_j}^{\prime\prime}   \circ {\cal M}_{ E_j}^{m-1}\;,   \nonumber \\
  \tilde{\cal C}_{{\cal S}, E_j}^{\prime\prime,b} &:= &
 \sum_{m' =m+1}^M  \sum_{m=1}^{M-1} {\cal M}_{ E_j}^{M-m'+1} \circ \tilde{\cal U}_{S_{m'},E_j}^{\prime}  
   \circ {\cal M}_{ E_j}^{m'-m}\circ \tilde{\cal U}_{S_{m},E_j}^{\prime}   \circ  {\cal M}_{ E_j}^{m-1}\;. \label{defCsec777}
 \end{eqnarray} 
As in the case of (\ref{EQperR111}) one can verify that the first order contribution nullifies. Indeed we have
\begin{eqnarray}
&&\!\!\!\! \!\!\!\! \!\!\!\! \!\!\!\! \!\!\!\! 
 \left\langle  \tilde{\cal C}^{\prime}_{{\cal S}, E_{n+1}}   \big({\bar{R}}(n)\otimes \eta\big) \right\rangle_{\cal E} 
:=-i \; \sum_{m}
  \left\langle   \left[
 {\Delta \hat{H}}_{S_m,E_{n+1}}   , {\bar{R}}(n) \otimes {\cal M}_{E_{n+1}}^{m-1}(\eta) ] \right]_- \right\rangle_{\cal E} \nonumber \\
&&=
 -i   \sum_{m}  \sum_{\ell}   \Big[  \hat{A}^{(n+1,\ell)}_{S_m} , {\bar{\rho}}(n) \Big]_-  \left\langle B^{(m,\ell)}_{E_{n+1}} {\cal M}_{ E_{n+1}}^{m-1} (\eta) \right\rangle_{E_{n+1}} =0\;,
 \end{eqnarray} 
because of Eq.~(\ref{IMPO}).
The remaining terms can be computed as in Eq.~(\ref{cprimeprimea}) and (\ref{cprimeprimeb}). Here we only stress on the fact that the component of $\tilde{\cal U}_{S_{m},E_j}^{\prime\prime}$
that depends upon the operator $\hat{Q}_{S_m,E_j}$ (i.e. the \emph{extra} term of Eq.~(\ref{NEWREVE2})) do not contribute in the final result. Indeed they only enters in the definition of $\tilde{\cal C}_{{\cal S}, E_j}^{\prime\prime,a}$ and produce
the following term
\begin{eqnarray}
&&\!\!\!\!  \!\!\!\!  \!\!\!\! 
-\frac{i}{2}  \; \sum_{m}
  \left\langle   \left[
\hat{Q}_{S_m,E_{n+1}}   , {\bar{R}}(n) \otimes {\cal M}_{E_{n+1}}^{m-1}(\eta) ] \right]_- \right\rangle_{\cal E}
\label{NEWIDENTITY} \\
&& \qquad =
\frac{g}{2}   \sum_{m} \left\langle  \Big[  \Big[ \hat{h}_{S_m}^{(n+1)},  {{\Delta \hat{H}}}_{S_m,E_{n+1}} \Big]_-,{\bar{R}}(n)\otimes  {\cal M}_{ E_{n+1}}^{m-1} (\eta) \Big]_-\right\rangle_{\cal E} 
\nonumber \\
&&\qquad=\frac{g}{2}   \sum_{m} \sum_{\ell} \left\langle  \Big[  \Big[ \hat{h}_{S_m}^{(n+1)},  \hat{A}^{(n+1,\ell)}_{S_m}  \otimes B^{(m,\ell)}_{E_{n+1}}
 \Big]_-,{\bar{R}}(n)\otimes  {\cal M}_{ E_{n+1}}^{m-1} (\eta) \Big]_-\right\rangle_{\cal E} \nonumber \\
&& \qquad  = \frac{g}{2}   \sum_{m} \sum_{\ell}   \Big[  \Big[ \hat{h}_{S_m}^{(n+1)},  \hat{A}^{(n+1,\ell)}_{S_m} 
 \Big]_-,{\bar{\rho}}(n)   \Big]_- \; \left\langle B^{(m,\ell)}_{E_{n+1}}
   {\cal M}_{ E_{n+1}}^{m-1} (\eta)\right\rangle_{E_{n+1}}  =0\;.\nonumber 
 \end{eqnarray} 
In summary Eq.~(\ref{exp1}) yields
 \begin{eqnarray}
\!\!\!\!\!\!\!\! \!\!\!\! \!\!\!\! \!\!\!\! 
 \frac{{\bar{\rho}}(n+1)-  {\bar{\rho}}(n)}{\Delta t}  =   \frac{g^2 \Delta t }{\gamma} \; \left\{  \sum_{m}  \bar{\cal L}_m ({\bar{\rho}}(n) )     
 + \sum_{m'>m} \bar{\cal D}^{(\rightarrow)}_{m,m'} ({\bar{\rho}}(n) )  \right\} 
\label{EQperR112456}   \;  + {\cal O}\big(g^3\Delta t^2\big) \;,
\end{eqnarray} 
 where now 
  \begin{eqnarray}
\!\!\!\! \!\!\!\! \!\!\!\!  \!\!\!\! \!\!\!\! \bar{\cal L}_m(\cdots) =  \frac{1}{2} \sum_{\ell,\ell'} 
{\gamma_m^{(\ell, \ell')}}
 \;  \Big[ 2 \bar{A}_{S_m} ^{(n+1,\ell')} (\cdots)\bar{A}_{S_m}^{(n+1,\ell)}  \nonumber \\
   -  \bar{A}^{(n+1,\ell)}_{S_m} \bar{A}^{(n+1,\ell')}_{S_m}(\cdots)-  (\cdots)\bar{A}^{(n+1,\ell)}_{S_m}  \bar{A}^{(n+1,\ell')}_{S_m}  \Big]   \;, \nonumber 
 \end{eqnarray}
 and 
  \begin{eqnarray}
\!\!\!\! \!\!\!\! \!\!\!\! \!\!\!\! 
 \bar{\cal D}^{(\rightarrow)}_{m,m'} (\cdots)& =&
  \sum_{\ell, \ell'}   \; \gamma^{(\ell,\ell')}_{m,m'}  \;  \bar{A}_{S_m}^{(n+1,\ell)}  \; \Big[ (\cdots) ,\bar{A}_{S_{m'}}^{(n+1,\ell')} \Big]_- \nonumber \\
&&-  \sum_{\ell, \ell'} \left[\gamma^{(\ell,\ell')}_{m,m'}\right]^* \;  \;   \Big[(\cdots), \bar{A}_{S_{m'}}^{(n+1,\ell')}   \Big]_- \;\bar{A}_{S_m}^{(n+1,\ell)}  
      \;, \nonumber 
 \end{eqnarray}
In these expressions the coefficients ${\gamma_m^{(\ell, \ell')}}$ and $\gamma^{(\ell,\ell')}_{m,m'}$ differ from those in 
Eqs.~(\ref{trieste3old})
and (\ref{gamma}) and are 
 expressed by
\begin{eqnarray}
{\gamma_m^{(\ell, \ell')}} &:=& \gamma \; \left\langle  B_E^{(m,\ell)}B_E^{(m,\ell')}  \; {\cal M}^{m-1}(\eta)\right\rangle_E\;, \nonumber \\
   \gamma^{(\ell,\ell')}_{m,m'}  &:=&   \gamma\; 
\left\langle B_{E}^{(m',\ell')} {\cal M}^{m'-m}\left(B^{(m,\ell)}_{E}  \; {\cal M}^{m-1}(\eta)\right) \; \right\rangle_{E}\;.
\end{eqnarray} 
Also the operators
$\bar{A}_{S_{m'}}^{(n+1,\ell')}$ stands for the operator ${A}_{S_{m'}}^{(n+1,\ell')}$ in the interaction representation~(\ref{DEFDEF})
induced by the transformation $V_{\cal S}^{(n)}$, i.e.
\begin{eqnarray}
 \bar{A}_{S_{m'}}^{(n+1,\ell')} :=  [ V_{\cal S}^{(n)}]^\dag \;  {A}_{S_{m'}}^{(n+1,\ell')}  \; V_{\cal S}^{(n)} \;.
\end{eqnarray} 
This follows from the fact that according to our definitions
\begin{eqnarray} 
\!\!\!\! 
\hat{\Theta}_{S_m,E_{n+1}} = [ V_{S_m}^{(n)}]^\dag  \Theta_{S_m,E_{n+1}}  V_{S_m}^{(n)}  =  [ V_{\cal S}^{(n)}]^\dag \;  \Theta_{S_m,E_{n+1}}  \; V_{\cal S}^{(n)} 
= : \bar{\Theta}_{S_m,E_{n+1}}\;.
\end{eqnarray} 
Taking now the limit~(\ref{defgamma}) this finally yields a differential equation for ${\bar{\rho}}(t)$ as in Eq.~(\ref{fin}) with $A_{S_m}^{(\ell)}$ operators being replaced by $\bar{A}^{(\ell)}_{S_m}(t):= \lim_{\Delta t\rightarrow 0^+} \bar{A}^{(n+1,\ell)}_{S_m}$.
It is worth noticing that in the continuos limit 
the transformation $V_{\cal S}^{(n)}$ which  define the interaction representation becomes:
 \begin{eqnarray}
V_{\cal S}^{(n)}&=&   \otimes_m \;  V_{S_m}^{(n)} =   \otimes_m e^{-i g  h_{S_m}^{(n)} \Delta t } \;e^{-i g  h_{S_m}^{(n-1)} \Delta t}  \cdots  e^{-i g  h_{S_m}^{(2)} \Delta t }\;  e^{-i g  h_{S_m}^{(1)} \Delta t }\nonumber \\
&=&\otimes_m {\cal T} \exp[ -i g \int_0^t h_{S_m}(t') dt']\;,
 \end{eqnarray}
with $h_{S_m}(t) := \lim_{\Delta t\rightarrow 0^+} h_{S_m}^{(n)}$.

\section{Conclusions} \label{conc}
In the present manuscript we have reviewed some of technical aspects of the new method recently introduced in Ref.~\cite{PRL} which allows
one to derive in a consistent way, general master equation for
cascade quantum system (i.e. multipartite quantum system which are unidirectionally coupled via a partially incoherent mediator). In particular
we focused on the main assumption of the model (the stability condition of Eq.~(\ref{ASSUMPTION1})) showing that it can be lifted by properly moving
into a interaction picture representation with respect to the free dynamics of the system of interest.

\end{document}